\documentclass[epj]{webofc}
\usepackage[varg]{txfonts}
\usepackage{hyperref}

\wocname{EPJ Web of Conferences}

\woctitle{CONF12}

\begin{document}

\selectlanguage{english}
\title{The quark propagator in QCD and $ G_2$ QCD}

\author{Romain Contant\inst{1}\fnsep\thanks{\email{romain.contant@uni-graz.at}} \and
        Markus Q. Huber\inst{1}\fnsep\thanks{\email{markus.huber@uni-graz.at}}}

\institute{Institute of Physics, University of Graz, NAWI Graz, Universitätsplatz 5, A-8010 Graz, Austria}

\abstract{
QCD-like theories provide testing grounds for truncations of functional equations at non-zero density, since comparisons with lattice results are possible due to the absence of the sign problem. As a first step towards such a comparison, we determine for QCD and $G_2$ QCD the chiral and confinement/deconfinement transitions from the quark propagator Dyson-Schwinger equation at zero chemical potential by calculating the chiral and dual chiral condensates, respectively.
}

\maketitle

\section{Introduction}
\label{intro}

Functional methods provide a non-perturbative approach to quantum field theory. These methods consist of infinitely large systems of (non-)linear equations. Thus, truncations and modeling are mandatory to solve them numerically. However, recent results indicate that a not too big number of correlation functions may be sufficient to achieve quantitatively reliable results \cite{Mitter:2014wpa,Cyrol:2016tym,Huber:2016tvc}. In general, lattice calculations provide a means to estimate the effects of truncations. However, at non-vanishing chemical potential quantum chromodynamics (QCD) suffers from the infamous sign problem \cite{deForcrand:2010ys}. On the other hand, some minimal modifications of QCD like changing the gauge group lead to theories with real and positive determinants for which lattice simulations at non-vanishing chemical potential are possible. Examples include QCD with the gauge group $SU(2)$ ($QC_{2}D$) with an even number of flavors \cite{Kogut:2000ek,Hands:2006ve} or with the gauge group $G_2$ ($G_{2}$ QCD) \cite{Holland:2003jy,Pepe:2006er,Maas:2012wr}. These simulations provide valuable information on medium effects and can be used as guides to built appropriate truncations of functional equations to investigate the phase diagrams of these theories. However, the applicability of the same truncation for different gauge groups is not clear yet.

In this study, we will compare the effect of temperature on the matter sector of QCD and a QCD-like theory with the gauge group $G_{2}$ at $\mu = 0 $. The objective is to take a first look at transitions for different gauges groups with a Dyson-Schwinger approach using the same truncation and modeling.
More precisely, we would like to know if the same truncation is sufficient to encode qualitatively the behavior for the chiral and confinement/deconfinement transitions for similar non-Abelian theories. The choice of $G_{2}$ is motivated by similarities between QCD and $G_2$ QCD. For example, both exhibit chiral and deconfinement transitions \cite{Holland:2003jy,Pepe:2006er,Cossu:2007dk,Danzer:2008bk}. A difference between $G_2$ and $SU(3)$ lies in their center, which is trivial for $G_2$. Nevertheless, the Polyakov loop can be used as order parameter. An advantage of $G_2$ QCD over $QC_2D$ is the presence of fermionic bound states. However, the spectrum of $G_2$ QCD is reacher than that of QCD \cite{Wellegehausen:2013cya}.

In this work we will compute the quark propagator in Landau gauge from its Dyson-Schwinger-equation (DSE). Temperature is incorporated through the Matsubara formalism. We employ a truncation scheme along the lines of \cite{Fischer:2012vc} using lattice results as input for the temperature dependence of the gluon dressing function and a model for the dressed quark-gluon vertex. As (pseudo-)order parameters for the quark confinement/deconfinement and chiral transitions, the dual quark condensate and the chiral condensate will be used. The details of the employed truncation and the setup will be discussed in Sect. \ref{sec-2}. Sect. \ref{sec-3} will be devoted to the numerical results of quenched and unquenched QCD and $G_{2}$ QCD. We conclude in Sec.~\ref{concl}.

\section{Gap equation, truncation and transitions}
\label{sec-2}

\subsection{Quark propagator and gap equation}
\label{sec-2a}

\begin{figure}[tb]
\centering
\includegraphics[width=11cm,clip]{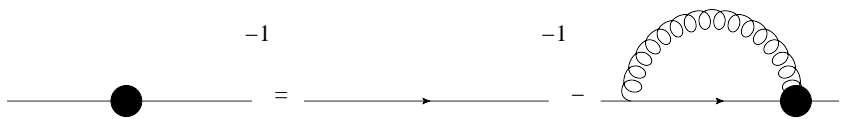}
\caption{The gap equation.
Quantities with a blob are fully dressed, as are internal propagators. Continuous/wiggly lines denote quarks/gluons.}
\label{fig-1}       
\end{figure}

In medium, the quark propagator can be written as
\begin{align}
S^{-1}(\vec{p}, \omega_{n}) = i \vec{p} \vec{\gamma} A(\vec{p}, \omega_{n}) + i \omega_{n} \gamma_{4} C(\vec{p}, \omega_{n}) + B(\vec{p}, \omega_{n}) +  i \omega_{n} \gamma_{4} \vec{p} \vec{\gamma} D(\vec{p}, \omega_{n}).
\end{align}
Its DSE is shown in fig. \ref{fig-1}. The self-energy reads
\begin{align}
\Sigma(\vec{p}, \omega_{n}) = Z_{1F} C_{F} (-g^{2})\sum_{q_4}{\int{\frac{d\vec{q}}{(2\pi)^3}\gamma_{\mu} S(q) \Gamma_{\nu}(p-q,q,p) D_{\mu \nu}(p-q)}}.
\end{align}
$D_{ \mu \nu }$ is the gluon propagator, $\Gamma^{\nu}_{q-gl}$ the dressed quark-gluon vertex, $Z_{1F}$ the quark-gluon vertex renormalization constant and $C_{F}$ the Casimir of the gauge group considered. Using the following projectors, the equations for individual dressing functions can be obtained:
\begin{align} 
P_{A} =& \frac{\vec{p} \vec{\gamma}}{4 i p^{2}},  P_{B} = \frac{1}{4},  P_{C} = \frac{\omega_{n}\gamma_{4}}{4i \omega^{2}_{n}},  P_{D} = \frac{\omega_{n} \gamma_{4} \vec{p} \vec{\gamma}}{4 i \omega^{2}_{n} \vec{p}^{2}}\\
A(p^{ 2 })&=Z_{ 2 }-\textrm{Tr}\left[P_{ A }\Sigma (p^{ 2 })\right], C(p^{ 2 })=Z_{ 2 }-\textrm{Tr}\left[P_{ C }\Sigma (p^{ 2 })\right],  \\
B(p^2) &= Z_{m} Z_{2} m - \textrm{Tr}\left[P_{B} \Sigma(p^2)\right], D(p^2) = - \textrm{Tr}\left[P_{D} \Sigma(p^2)\right]. \end{align}
$Z_{2}$, $Z_{m}$ and $m$ are the renormalization constants for the quark wave function, the renormalization constant for the quark mass and the bare quark mass.

\subsection{Gluon input}
\label{sec-2b}

At finite temperature, the gluon propagator can be decomposed into two parts,\begin{align}
D_{\mu \nu}(p) = P^{L}_{\mu \nu}(p) \frac{Z^{L}(p^2)}{p^2} + P^{T}_{\mu \nu}(p) \frac{Z^{T}(p^2)}{p^2}, \end{align}
with $Z_{L}$ and $Z_T$ the gluon dressing functions longitudinal and transverse with respect to the heat bath, respectively. 
The corresponding results from gauge-fixed lattice simulations \cite{Fischer:2010fx,Maas:2011ez} can be fitted to~\cite{Fischer:2010fx}
\begin{align}
Z_{ T/L }(x)=\frac { x }{ (x+1)^2 } \left( \left( \frac { c }{ x+a_{ T/L } }  \right) ^{ b_{ T/L } }+x\left( \frac { \alpha (\mu )\beta _{ 0 } }{ 4\pi  } \textrm{ln}(x+1)) \right) ^{ \gamma  } \right),\end{align}
where $x = \frac{p^2}{\Lambda^{2}} $ , $\gamma = \frac{-13 C_A + 4 N_f}{22 C_A - 4 N_f}$ is the anomalous dimension of the gluon and $\beta_{0} = \frac{11 C_{A} - 2 N_{f}}{3}$ with $C_A$ the Casimir in the adjoint representation of the gauge group considered. $\Lambda^{2}$ and $c$ are temperature independent parameters, while $a_{T/L}$ and $b_{T/L}$ are the temperature dependent fitting parameters. To interpolate between the available temperatures, one can fit the temperature dependence of $a_{T/L}$ and $b_{T/L}$. We use the fit given in \cite{Luecker13}. Only the first Matsubara mode is considered in the fit, and higher modes are accessed by $Z^{T/L}(\vec{p}^2, \omega_{n}) \rightarrow Z^{T/L}(\vec{p}^2 + \omega_{n}^{2}, 0)$. For $G_2$, the gluon dressing functions are to our knowledge not available from lattice calculations in four dimensions at non-zero temperature. However, results in three dimensions show a good agreement between the gluon propagators of $SU(3)$ and $G_2$ for zero temperature \cite{Maas:2007af}. This motivates using the $SU(3)$ fits for $G_2$, but with the temperature rescaled to match the critical temperature of $G_2$. We compensate the change in the UV part of the fit induced by the different value of $\beta_0$ for $G_2$ by changing $\alpha(\mu)$ accordingly.

\subsection{Quark-gluon vertex model}
\label{sec-2c}

At finite temperature, we do not have much information about the dressed quark-gluon vertex neither from lattice calculations nor from functional methods. Thus, for now we have to rely on models.
The model employed in the following is given by \cite{Fischer:2009wc}
\begin{align}
\Gamma_{\nu}(q,p,l) &= \gamma_{\mu} \Gamma_{mod}(x)\left(\frac{A(p^2) + A(l^2)}{2} \delta_{\mu, i}+\frac{C(p^2) + C(l^2)}{2} \delta_{\mu, 4} \right).
\\
\Gamma_{mod}(x) &= \frac{d_{1}}{\left(x+d_2\right)} + \frac{x}{\Lambda^2 + x} \left(\frac{\alpha(\mu)\beta_{0}}{4 \pi}\textrm{ln}\left(\frac{x}{\Lambda^2} + 1\right)\right)^{2 \delta}.
\end{align}
$p$ and $l$ are the quark and anti-quark momenta and $q$ is the gluon momentum. For technical reasons of renormalizability, the choice for $x$ depends on the equation in which the vertex model is used. In the gluon propagator DSE it is $(p^2+l^2)$ and in the quark propagator DSE $q^2$. $\delta$ is the anomalous dimension of the ghost for which $\gamma + 2 \delta = 1 $ holds. This model contains only the tree-level tensor. Since it is known that also other dressing functions of the quark-gluon vertex become important \cite{Hopfer:2013np,Mitter:2014wpa,Williams:2015cvx}, it is attempted to effectively capture their contributions in the IR part where the parameters $d_{1}$ and $d_{2}$ are temperature independent for now.

\subsection{Transitions}
\label{sec-2d}

We will determine the transitions from the chiral and dual chiral condensates. 
The former is calculated from the quark propagator as 
\begin{align} \left<  \overline{\psi} \psi \right>  = -C_A Z_2 Z_m T \sum_{l_4}{\int{\frac{d\vec{l}}{(2\pi)^3}\textrm{Tr}[S(l)]}}. \end{align}
A non-zero value means that chiral symmetry is broken.
For non-zero bare quark masses, there is no exact chiral symmetry. The chiral condensate is then non-zero above the transition but still small compared to the low temperature phase. The UV divergence of the chiral condensate is renormalized by subtracting a quark condensate with a heavier bare mass from a condensate with a lighter bare mass: $\Delta_{l,h} = -\left<  \overline{\psi} \psi \right>_{l} + \frac{m_l}{m_s}\left<  \overline{\psi} \psi \right>_{h}$
where $m_l$ and $m_s$ are the bare masses for a lighter and a heavier quark, respectively, and the sign was chosen such as to make $\Delta_{l,h}$ positive.

The second transition studied here is the confinement/deconfinement transition. The typical order parameter is the Polyakov loop which is related to the center symmetry of QCD. The dual quark condensate \cite{Bilgici:2008qy,Synatschke:2007bz} is proportional to the Polyakov loop and can serve as an alternative order parameter that is accessible with functional methods \cite{Fischer:2009wc}.
To compute the dual quark condensate $\Sigma$, we introduce the generalized $U(1)$ valued boundary condition $\psi(x, 1/T) = e^{i \phi } \psi(x, 0)$ where the physical condition is given by $\phi = \pi$.
In a lattice formulation, the generalized condensate corresponds to a sum over closed loops winding $n$ times around the temporal direction. One can project this quantity to $n=1$ with a Fourier transformation.

\section{Results }
\label{sec-3}

\begin{figure}[tb]
   \begin{minipage}[tb]{0.40\linewidth}
      \centering \includegraphics[width=6.5cm,clip]{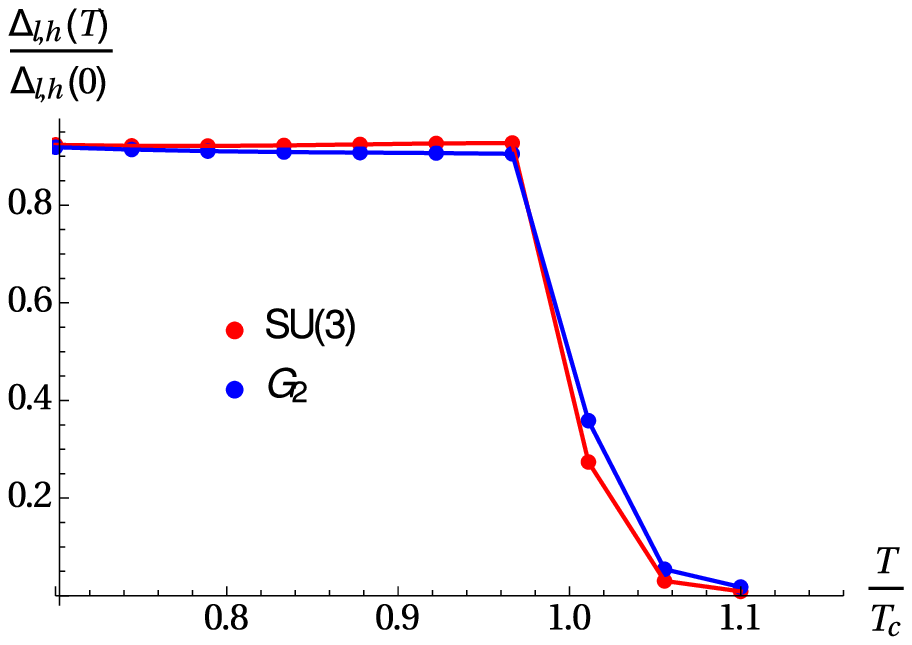}
      \caption{Temperature evolution of the chiral condensate for quenched $SU(3)$ and $G_2$. The chiral condensate is normalized by its value in vacuum.}
      \label{fig-2}
   \end{minipage}\hfill
   \begin{minipage}[tb]{0.46\linewidth}
	\centering \includegraphics[width=6.5cm,clip]{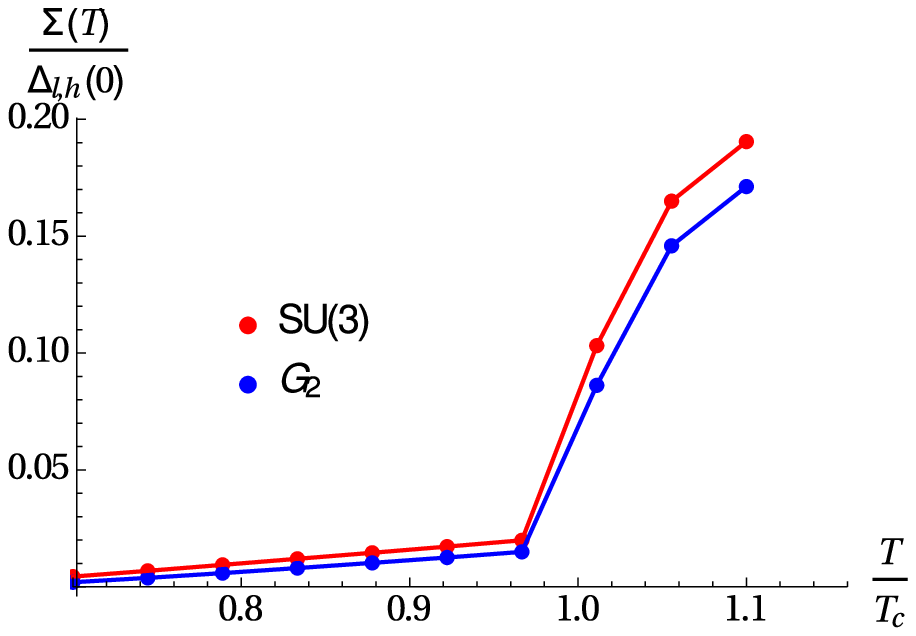}
    \caption{Temperature evolution of the quenched dual condensate for $SU(3)$ and $G_2$. The dual condensate is normalized by the value of the chiral condensate in vacuum.}
	\label{fig-3}       
   \end{minipage}\hfill
     
\end{figure}

We first compare $SU(3)$ and $G_2$ in the quenched case. The different parameters are summarized in table \ref{tab-1}. The results for the chiral and dual quark condensates are show in figs.\ref{fig-2} and \ref{fig-3}, respectively. The quark condensate is normalized by its vacuum value. For the confinement/deconfinement transition, $SU(3)$ and $G_2$ show a rapid increase of the dual quark condensate after the critical temperature. The non-zero value of the dual quark condensate at low temperatures is most likely due to some sensitivity to the parameters of the employed input which was already observed previously \cite{Fischer:2009wc,Luecker13}.

\begin{table}[tb]
\begin{minipage}{0.50\textwidth}
\caption{The critical temperatures for the quenched computations and employed parameters. $\mu$ is the renormalization point for the quark DSE.\smallbreak $\Lambda=1.4\,\text{GeV}$, $c=11.5\,\text{GeV}^2$, $d_2$ = 0.5 MeV,  $m( \mu = 80 \text{GeV} )$ = 3 MeV, $N_f$ = 0}
\label{tab-1}
\begin{tabular}{lll}
\hline
       & $SU(3)$ & $G_2$   \\\hline \hline
        \multicolumn{2}{l}{Parameters} &  \\\hline
       $d_1$ & 4.5 MeV & 5.6 MeV \\\hline
        \multicolumn{2}{l}{Critical temperature}&  \\\hline
$T_c$ & 277 MeV \cite{Fischer:2010fx} & 255 MeV \cite{Ilgenfritz:2012wg,Cossu:2007dk} \\
\end{tabular}
\end{minipage}
\hfill
\begin{minipage}{0.50\textwidth}
\caption{The critical temperatures for the unquenched computations. \smallbreak $\Lambda=1.4\,\text{GeV}$, $c=11.5\,\text{GeV}^2$, $d_2$ = 0.5 MeV,  $m( \mu = 80 \text{GeV} )$ = 1.2   MeV, $N_f$ = 2}
\label{tab-2}
\begin{tabular}{lll}
\hline
       & $SU(3)$ & $G_2$   \\\hline \hline
        Parameters &  \\\hline
       $d_1$ & 7.6 MeV & 6.2 MeV \\
       $\alpha(\mu)$ & 0.3 & 0.48 \\\hline
        \multicolumn{2}{l}{Critical temperature}& \\\hline
$T_c$ (chiral) & 204 MeV & 156 MeV \\
$T_c$ (deconfinement) & 205 MeV & 164 MeV \\
\end{tabular}
\end{minipage}
\end{table}

\begin{figure}[tb]
\centering
\includegraphics[width=11cm,clip]{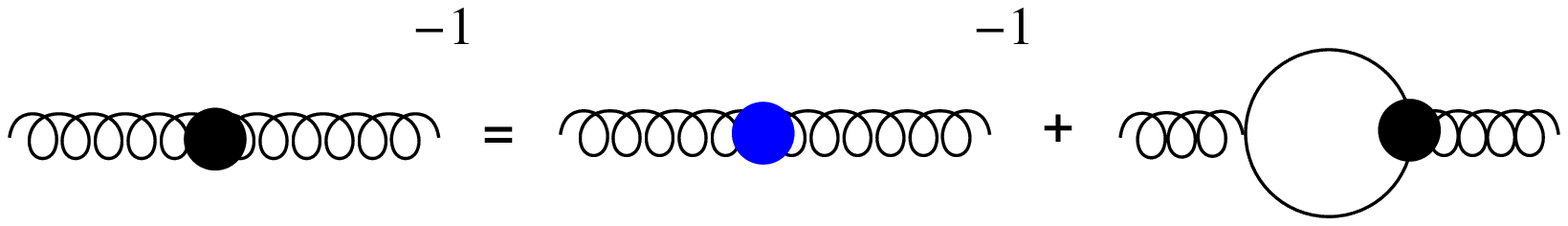}
\caption{Approximated DSE for the gluon propagator. The blue dot represents the quenched gluon input fitted from lattice data.}
\label{fig-4}       
\end{figure}
\begin{figure}[tb]
   \begin{minipage}[tb]{0.40\linewidth}
      \centering \includegraphics[width=6.5cm,clip]{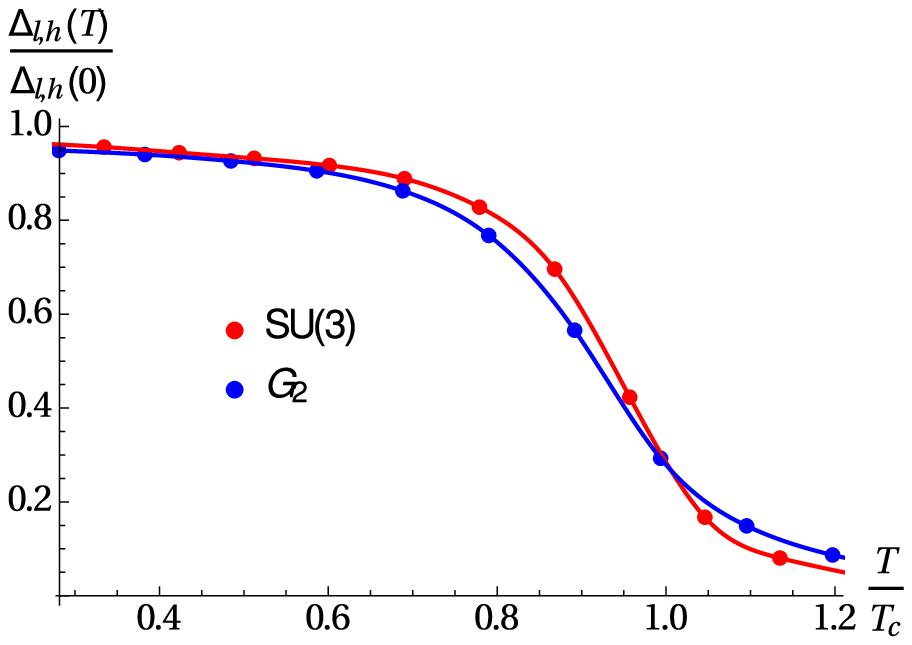}
      \caption{Temperature evolution of the unquenched chiral condensate for $SU(3)$ and $G_2$. The chiral condensate is divided by its value in vacuum. }
      \label{fig-5}
   \end{minipage}\hfill
      \begin{minipage}[tb]{0.46\linewidth}
      \centering \includegraphics[width=6.5cm,clip]{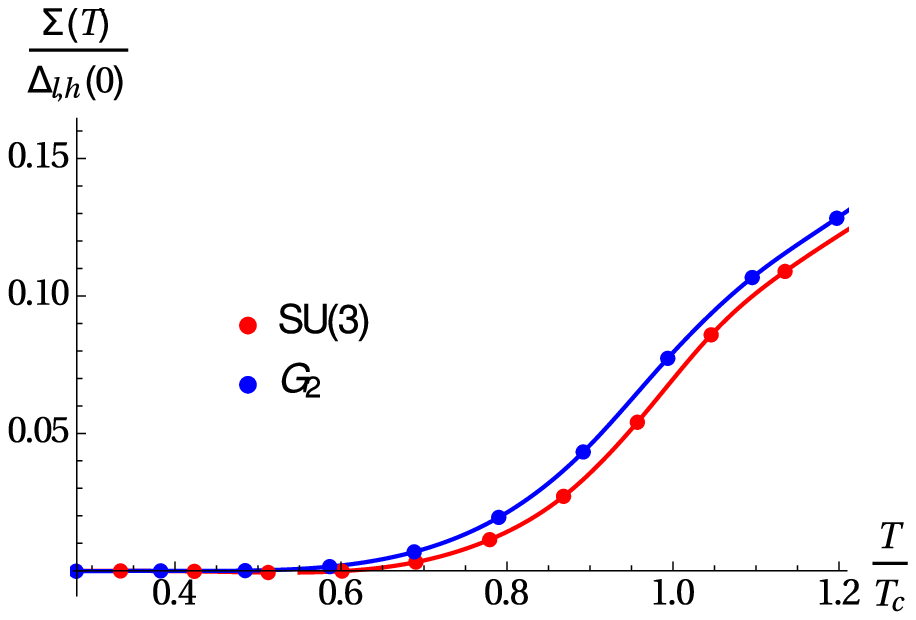}
      \caption{Temperature evolution of the unquenched dual condensate for $SU(3)$ and $G_2$. The dual condensate is divided by the value of the chiral condensate in vacuum.}
      \label{fig-6}
   \end{minipage}\hfill
\end{figure}

Adding dynamical quarks requires to solve the coupled system of quark, gluon and ghost propagator DSEs. However, instead of solving the Yang-Mills part we approximate it by using the quenched lattice results and calculate only the quark-loop dynamically \cite{Fischer:2012vc}, see fig.~\ref{fig-4}. This adds all direct quark contributions but not indirect ones which would enter via the unquenched gluon propagator in the Yang-Mills part.
To discard spurious divergences, we use a generalized Brown-Pennington projector \cite{Fischer:2012vc}.
The parameter $d_1$ is fixed to obtain the same value of $f_{\pi}$ in vacuum for $G_2$ and $SU(3)$.

For $p^2 \rightarrow 0$,  $\Pi^{L}$ will be divergent as $\Pi^{L} p^2 \rightarrow 2 m_{th}^2$, where $m_{th}$ is the Debye mass. In the unquenched case, the phase transitions become crossovers. Their critical temperatures are defined as the extrema of the derivatives of the pseudo-order parameters.
Tab.~\ref{tab-2} summarizes the obtained critical temperatures as well as the used model parameters. The results of the computation are shown in fig.\ref{fig-5} for the chiral condensates and in fig. \ref{fig-6} for the dual condensates.

The chiral transition temperature is reduced by approximately $25\,\%$ for $SU(3)$ as compared to the quenched case and by almost $40\,\%$ for $G_2$. The confinement/deconfinement and chiral transitions occur approximately at the same place for $SU(3)$ while they are a bit shifted for $G_2$, see tab.~\ref{tab-2}. Finally, the value of the dual quark condensate is always greater for $G_2$ than for $SU(3)$.
In general, the $G_2$ results look very similar to the QCD results.

\section{Conclusions}
\label{concl}

This study was devoted to a first comparison of QCD and $G_2$ QCD within the DSE formalism. For $G_2$ we employed a truncation in analogy to the $SU(3)$ case \cite{Fischer:2012vc}. Due to the lack of lattice data for the quenched gluon propagator required as input, we used the $SU(3)$ fits with the temperature rescaled to match the critical temperature of quenched $G_2$. For the unquenched calculations we find good qualitative and even quantitative agreement between the two cases. 

Further work will be required to shed more light on the extent to which truncations for $G_2$ and $SU(3)$ can be constructed analogously. Discarding the model input would mean to calculate not only all propagators dynamically, but in particular calculating the quark-gluon vertex. On the other hand, the employed approximation can be directly generalized to non-vanishing chemical potential, where detailed results for QCD \cite{Fischer:2012vc} are available. In another direction, calculations of two-color QCD would provide an additional angle at the question of the gauge group dependence of truncations.

\section{Acknowledgments}

\label{Ackno}
The numerical computation was performed with CrasyDSE \cite{Huber:2011xc} using HPC Clusters at the University of Graz. Feynman diagrams were created with Jaxodraw \cite{Binosi:2003yf}.
Support by the FWF (Austrian science fund) under Contract No.P27380-N27 is gratefully acknowledged. 

\bibliography{lit_CONF12}

\begin{thebibliography}{25}

\bibitem{Mitter:2014wpa}
M.~Mitter, J.M. Pawlowski, N.~Strodthoff, Phys. Rev. \textbf{D91}, 054035
  (2015), \texttt{1411.7978}

\bibitem{Cyrol:2016tym}
A.K. Cyrol, L.~Fister, M.~Mitter, J.M. Pawlowski, N.~Strodthoff, Phys. Rev.
  \textbf{D94}, 054005 (2016), \texttt{1605.01856}

\bibitem{Huber:2016tvc}
M.Q. Huber, Phys. Rev. \textbf{D93}, 085033 (2016), \texttt{1602.02038}

\bibitem{deForcrand:2010ys}
P.~de~Forcrand, PoS \textbf{LAT2009}, 010 (2009), \texttt{1005.0539}

\bibitem{Kogut:2000ek}
J.B. Kogut, M.A. Stephanov, D.~Toublan, J.J.M. Verbaarschot, A.~Zhitnitsky,
  Nucl. Phys. \textbf{B582}, 477 (2000), \texttt{hep-ph/0001171}

\bibitem{Hands:2006ve}
S.~Hands, S.~Kim, J.I. Skullerud, Eur.Phys.J. \textbf{C48}, 193 (2006),
  \texttt{hep-lat/0604004}

\bibitem{Holland:2003jy}
K.~Holland, P.~Minkowski, M.~Pepe, U.J. Wiese, Nucl. Phys. \textbf{B668}, 207
  (2003), \texttt{hep-lat/0302023}

\bibitem{Pepe:2006er}
M.~Pepe, U.J. Wiese, Nucl. Phys. \textbf{B768}, 21 (2007),
  \texttt{hep-lat/0610076}

\bibitem{Maas:2012wr}
A.~Maas, L.~von Smekal, B.~Wellegehausen, A.~Wipf, Phys. Rev. \textbf{D86},
  111901 (2012), \texttt{1203.5653}

\bibitem{Cossu:2007dk}
G.~Cossu, M.~D'Elia, A.~Di~Giacomo, B.~Lucini, C.~Pica, JHEP \textbf{10}, 100
  (2007), \texttt{0709.0669}

\bibitem{Danzer:2008bk}
J.~Danzer, C.~Gattringer, A.~Maas, JHEP \textbf{01}, 024 (2009),
  \texttt{0810.3973}

\bibitem{Wellegehausen:2013cya}
B.H. Wellegehausen, A.~Maas, A.~Wipf, L.~von Smekal, Phys. Rev. \textbf{D89},
  056007 (2014), \texttt{1312.5579}

\bibitem{Fischer:2012vc}
C.S. Fischer, J.~Luecker, Phys.Lett. \textbf{B718}, 1036 (2013),
  \texttt{1206.5191}

\bibitem{Fischer:2010fx}
C.S. Fischer, A.~Maas, J.A. Muller, Eur. Phys. J. \textbf{C68}, 165 (2010),
  \texttt{1003.1960}

\bibitem{Maas:2011ez}
A.~Maas, J.M. Pawlowski, L.~von Smekal, D.~Spielmann, Phys.Rev. \textbf{D85},
  034037 (2012), \texttt{1110.6340}

\bibitem{Luecker13}
J.~Luecker, Ph.D. thesis, Justus-Liebig-Universit{\"a}t (2013),
  \texttt{http://geb.uni-giessen.de/geb/volltexte/2013/10128}

\bibitem{Maas:2007af}
A.~Maas, S.~Olejnik, JHEP \textbf{02}, 070 (2008), \texttt{0711.1451}

\bibitem{Fischer:2009wc}
C.S. Fischer, Phys. Rev. Lett. \textbf{103}, 052003 (2009), \texttt{0904.2700}

\bibitem{Hopfer:2013np}
M.~Hopfer, A.~Windisch, R.~Alkofer, PoS \textbf{ConfinementX}, 073 (2012),
  \texttt{1301.3672}

\bibitem{Williams:2015cvx}
R.~Williams, C.S. Fischer, W.~Heupel, Phys. Rev. \textbf{D93}, 034026 (2016),
  \texttt{1512.00455}

\bibitem{Bilgici:2008qy}
E.~Bilgici, F.~Bruckmann, C.~Gattringer, C.~Hagen, Phys. Rev. \textbf{D77},
  094007 (2008), \texttt{0801.4051}

\bibitem{Synatschke:2007bz}
F.~Synatschke, A.~Wipf, C.~Wozar, Phys. Rev. \textbf{D75}, 114003 (2007),
  \texttt{hep-lat/0703018}

\bibitem{Ilgenfritz:2012wg}
E.M. Ilgenfritz, A.~Maas, Phys. Rev. \textbf{D86}, 114508 (2012),
  \texttt{1210.5963}

\bibitem{Huber:2011xc}
M.Q. Huber, M.~Mitter, Comput.Phys.Commun. \textbf{183}, 2441 (2012),
  \texttt{1112.5622}

\bibitem{Binosi:2003yf}
D.~Binosi, L.~Theussl, Comput.Phys.Commun. \textbf{161}, 76 (2004),
  \texttt{hep-ph/0309015}

\end{thebibliography}

\end{document}